\documentclass[preprints,article,accept,moreauthors,pdftex,a4paper]{Definitions/mdpi} 
\pdfoutput=1
\usepackage{pdflscape}
\usepackage{wrapfig}
\usepackage{lscape}
\usepackage{rotating}
\usepackage{enumitem}
\usepackage{pgfplots}
\usepackage{graphicx}
\pgfplotsset{compat=1.15}
\usetikzlibrary{arrows.meta}
\usepackage[edges]{forest}
\forestset{
  declare dimen register=gap,
  gap'=5mm,
  declare count register=twist,
  twist'=1,
  family tree too/.style={
    draw,
    forked edges,
    for tree={
      fill=gray!0,
      rounded corners=1pt,
      minimum width/.wrap pgfmath arg={##1}{(\textwidth-6*(gap))/3},
      minimum height=4ex,
      font=\sffamily,
      text centered,
      edge=thick,
    },
    where={level()<(twist)}{%
      parent anchor=children,
      l sep+=5pt,
    }{%
      draw,
      folder,
      grow'=0,
      if={level()==(twist)}{%
        before typesetting nodes={child anchor=north},
        anchor=north,
        edge path'={%
          (!u.parent anchor) -- (.child anchor)
        },
      }{
        l sep'+=2pt,
        if level=2{
          before computing xy={
            l/.wrap pgfmath arg={##1}{14pt-((\textwidth-6*(gap))/6)}
          },
        }{}
      },
    },
  },
}
\firstpage{1} 
\pubvolume{11}
\issuenum{15}
\articlenumber{6790}
\pubyear{2021}
\copyrightyear{2021}
\externaleditor{{Academic Editor}:Luigi Pomante } 
\datereceived{25 June 2021} 
\dateaccepted{21 July 2021} 
\datepublished{21 July 2021} 
\hreflink{https://doi.org/10.3390/app11156790} 

\hypersetup{
pdftitle={Physical Side-Channel Attacks on Embedded Neural Networks: A Survey},
pdfsubject={cs.CR, cs.LG, cs.NE},
pdfauthor={Maria M\'endez Real, Rub\'en Salvador},
pdfkeywords={Physical Side-Channel Attacks, Side-Channel Analysis, Hardware Security, Machine Learning, Embedded Neural Networks, Deep Neural Networks, Security of Neural Networks},
}

\Title{Physical Side-Channel Attacks on Embedded Neural Networks: A Survey}

\TitleCitation{Physical Side-Channel Attacks on Embedded Neural Networks: A Survey}

\Author{Maria M\'endez Real $^{1,}$*\orcidA{} and Rub\'en Salvador $^{2}$\orcidB{}}

\AuthorNames{Maria M\'endez Real and Rub\'en Salvador}

\AuthorCitation{M\'endez Real, M.; Salvador, R.}

\address{%
$^{1}$ \quad Univ Nantes, CNRS, IETR UMR 6164, F-44000 Nantes, France\\
$^{2}$ \quad CentraleSup\'elec, CNRS, IETR UMR 6164, F-35576 Cesson-S\'evign\'e, France; ruben.salvador@centralesupelec.fr}

\corres{Correspondence: maria.mendez@univ-nantes.com}

\abstract{During the last decade, Deep Neural Networks (DNN) have progressively been integrated on all types of platforms, from data centers to embedded systems including low-power processors and, recently, FPGAs. 
{Neural Networks (NN)} are expected to become ubiquitous in IoT systems by transforming all sorts of real-world applications, including applications in the safety-critical and security-sensitive domains. 
However, the underlying hardware security vulnerabilities of embedded NN implementations remain unaddressed. In particular, embedded {DNN} implementations are vulnerable to Side-Channel {Analysis (SCA) attacks}, which are especially important in the IoT {and edge computing} context{s} where an attacker can {usually} gain physical access to the targeted device.
A research field has therefore emerged and is rapidly growing in terms of the use of SCA including timing, electromagnetic attacks and power attacks to target NN embedded implementations. Since 2018, research papers have shown that SCA enables an attacker to recover inference models architectures and parameters, to expose industrial IP and endangers data confidentiality and privacy. Without a complete review of this emerging field in the literature so far, this paper {surveys} state-of-the-art physical SCA {attacks relative to the implementation of} embedded {D}NN{s on micro-controllers and FPGAs} in order to provide a thorough analysis on the current {landscape}. It provides a taxonomy and a detailed classification of current attacks. It first discusses mitigation techniques and then provides insights for future research leads.}


\begin{document}
\section{Introduction}

Every electronic device generates {observable} parasite signals during {(i.e., as an unintentional side effect of)} data computation {that can leak internal information}. {The sources of this information leakage, which are known as side-channels,} can be {computation} timing, power consumption or electromagnetic {(EM)} emanation among others. {Physical side-channels depend on the transistor switching activity of a computing device. Therefore, they} are dependent on processing internal states and, {hence,} on the manipulated data. By statistically analysing {the measured side-channel data} for each internal state together with a hypothesis on the processed data, an attacker can deduce some intermediate {computation} states{. This in turn allows the extraction of} information {about} the processed data that would be otherwise inaccessible. {SCA} is a very well known research field in security and crypt analysis where it has enabled the recovery of secret keys in cryptographic algorithms,which are otherwise considered mathematically secure. 

In the same manner {as} cryptographic implementations, Machine Learning (ML) implementations are vulnerable to {SCA attacks}. This work especially focuses on Deep Neural Networks (DNN), as they are a family of ML algorithms extensively used in real-world industrial applications including applications in safety-critical and security-sensitive domains, such as autonomous and intelligent driving, healthcare, smart manufacturing, security and defense among others. {DNNs} are especially used for recognition, classification and analysis tasks.

There is increasing interest in deploying {DNNs} on low-cost {embedded} processors and {accelerators such as General Purpose Graphics Processing Units (GPU) and,} lately, also on Field Programmable Gate Arrays (FPGA)~\cite{mittal_SurveyFPGAbasedAccelerators_2020}. {FPGAs} have proven their efficiency for energy constrained {and highly customised DNN} implementations, especially for inference and small size networks~\cite{abdelouahab2018accelerating,chen2016eyeriss,abdelouahab2017tactics}. {DNNs} are expected to become ubiquitous in IoT systems in the next years.

In this context, physical {SCA attacks} are a real threat since attackers can easily gain physical access to the targeted device. A new and rapidly evolving research field has emerged around the exploitation of SCA on embedded {D}NN {implementations}, showing that it is possible to expose industrial secrets such as {D}NN model architecture and parameters as well as to endanger data confidentiality and privacy exposing training and/or input data. 
If recent work studies the possibility of deep learning computation over encrypted data (encrypted data and/or encrypted parameters)~\cite{qaisarahmadalbadawi2020towards,bourse2018fast,hesamifard2017cryptodl,Chillotti}, i.e., by homomorphic encryption, today's deep learning computation manipulates non{-}encrypted data.

Therefore, the study of SCA {attacks} against embedded {DNN} implementations is of great importance. It is indeed necessary to better understand and categorise literature attacks of today's classical {D}NN implementations in order to propose adapted and efficient countermeasures. Note that, to the best of our knowledge, no attack has been performed against {side-channel} protected {D}NN implementations. 
Without a complete review in the literature so far, this work aims {at} providing a thorough analysis and classification of state-of-the-art physical passive attacks on embedded {D}NN implementations. After proposing a taxonomy of SCA {attacks}, {we} provide a detailed classification of current attacks according to several features such as target model and {D}NN implementation, source of leakage, aim of the attack and the considered threat model. Finally, {we} discuss {the} first {reported} mitigation techniques. The main objective {and contribution} of this paper is to provide a broad overview of SCA {attacks} against embedded {D}NN implementations {on micro-controllers and FPGAs} for readers who wish to start working on this topic as well as to provide insights into the possible mitigation solutions and future research leads.

The remainder of this paper is organised as follows. Section~\ref{sec:Background} provides some background on {D}NN{s}, SCA in general and {on the specific} SC{A} techniques that will be covered in the surveyed attacks. Then, Section~\ref{sec:Threat_model_and_motivation} presents the considered threat model and motivates SCA {attacks} against {D}NN implementations. Section~\ref{sec:Physical_SCA_on_NN} first proposes a taxonomy of {these} SCA {attacks}, then it discusses literature works and classifies them according to several different features. Section~\ref{sec:Current_countermeasures} presents {the} first countermeasures proposed so far and highlights their limitations. Section~\ref{sec:Discussion_and_future_research_leads} discusses {existing} challenges and provides future research directions. Finally, Section~\ref{sec:Summary} concludes this work.

\section{Background}
\label{sec:Background}

\subsection{Some Background on {Neural Networks (NN)}}

{The existing attacks on the} literature have mainly targeted two classes of NN{s} (in the remainder of this manuscript, we use the acronym NN to refer generically to any type of neural network, irrespective of whether the network is actually deep or not. We use the terms DNN/CNN whenever required to be more specific) mostly used for classification problems: Multi Layer Perceptron {(MLP)} and Convolution{al} Neural Networks {CNN}.
This section {provides} some background on NNs {as} related to the attacks surveyed in this paper{. T}he reader is referred to~\cite{Goodfellow-et-al-2016} for {more} detailed information with respect to {D}NN{s}.

\begin{figure}[H]
\resizebox{.9\linewidth}{!}{\includegraphics{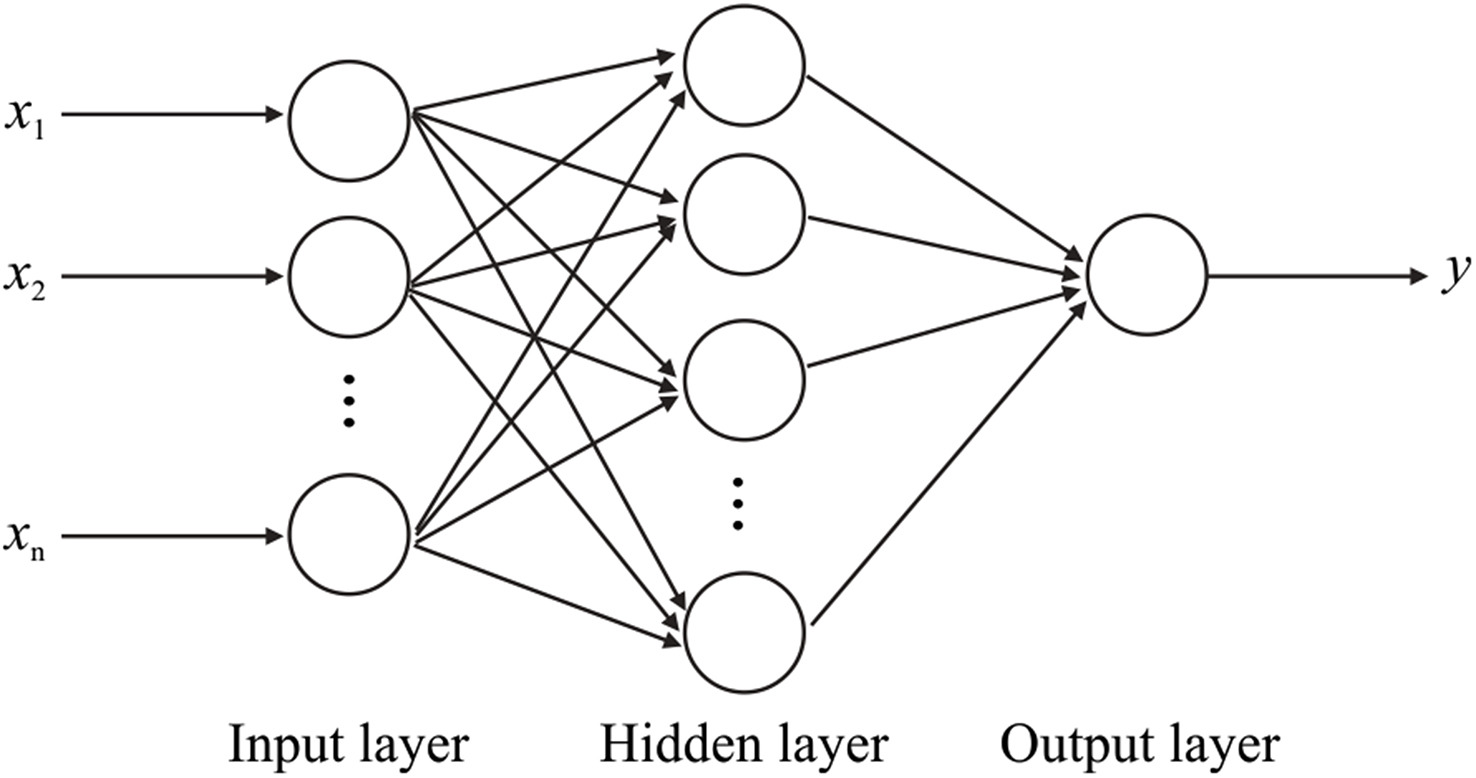}}
\caption{Multi Layer Perceptron, where $x$ is the input and $y$ is the output of a neuron.}
\label{MLP}
\end{figure}

{An} MLP is a feed-forward network which {is composed of several layers and each layer comprises} a number of neurons.
Each neuron of each layer is connected with a certain strength connection called weight to every node of the next layer. Data circulates in one manner, from the input layer to the output layer (feed-forward). MLP{s} include at least three different layers{:} an input layer, an output layer and one or more hidden layers (see \figurename~\ref{MLP}). 

Each neuron receives as inputs ($x$) the results of its previous layer connection(s), calculates a weighted summation added with a certain \textit{bias} (see Equation~(\ref{output_neuron})) and then applies a nonlinear transformation to finally compute its activation output ($y$). This latter output is passed onto the {next }connected layer neurons as their input. 

\begin{equation}
y = Activation \left(\sum \left(weight \cdot x \right) + bias\right)
\label{output_neuron}
\end{equation}

Different nonlinear \textit{activation} functions exist, including logistic (sigmoid), Tanh, softmax and Rectified Linear Unit (ReLU) functions. ReLU is one of the most widely used one{s} as it has the particularity of activating only a certain number of neurons at the same time, rather tha{n} activating them all at a given moment. The network is then sparse and easier to compute as only a subset of neurons is activated at any time~\cite{sharma2017activation}.

CNN{s} are a variant of the standard MLP NN{s} mainly used for the classification of multi-dimensional data, {e}specially images. CNN{s} are distinguished by the convolution layer. The latter is responsible for the feature extraction through a convolution operation between the high-dimensional input data and kernels, which are small matrices of parameters, in order to generate the layer output. Finally, each output layer is passed as input to the next layer. In addition to {the} convolution layer, other layers might be included in CNN{s} such as non-linear functions, pooling and fully-connected layers. {Further} details on CNN{s} can be found in~\cite{sze2017efficient,Goodfellow-et-al-2016}.

BNN{s} (Binarized Neural Networks)~\cite{hubara2016binarized} are a subfamily of CNN{s} that binarize {input and output} activation{s} and weights values. This converts floating point multiplication operations {in}to single-bit XNOR operation{s}, which significantly reduce storage requirements and computational complexity. 
However, binarization cause{s} a loss of information and deviation from full-precision results; therefore, {optimized training strategies are} still being studied~\cite{qin2020binary}. 
BNNs are a promising solution for IoT/edge resource{-}constrained and power{-}constrained devices where trad{ing} some accuracy loss for efficiency {can be accepted}.

All NN{s} have two stages: training and inference. In the training stage, the network is fed with pre-classified {training} inputs {along} with their classification results{, eventually converging} into {a set of} parameters values (weights and bias) able to compute the correct output results for its specific classification problem. During the inference stage{,} the pre-trained network can perform its task by processing the input data and computing the output by using the weights determined during the training stage. 

When looking at {a standard} NN model, we can refer to its architecture consisted of \textit{macro-parameters}, such as number of layers, neurons per layer and activation function, and to its \textit{micro-parameters} (or {simply} \textit{parameters}) such as weights and biases.

{The} attacks studied in this paper have targeted MLP, CNN and BNN models. However, other ML algorithms such as decision trees have also been targeted by physical SCA {attacks}~\cite{jap2020practical}.

\subsection{{SCA} Generalities}
\label{subsec:Generalities}
Every physical implementation {(of a processing system based on current silicon technologies)} generates unintentional {side-channel emissions} that can be observed and analysed. While {the system is} manipulating data, an attacker can observe these {side-channel emissions}, e.g., power consumption, EM, {computation} time, memory access patterns, etc. Then, the attacker investigates the dependencies of the observed {side-channel} signatures and the hypothetical signatures issued of the prediction of intermediate data and computation states. {This process can eventually disclose internal data from the system and effectively leaks private information.}

SC{A} is a well studied research field and has traditionally been used to attack cryptograph{ic} implementations of theoretically secure algorithms in order to recover the secret key. 
However, SCA {attacks} are more general and have been successfully applied to other targets in different scenarios{,} including NNs.

\begin{figure}[H]
\resizebox{.36\linewidth}{!}{
\begin{minipage}{0.36\textwidth}
    \begin{forest}
        family tree too,
        [Side-Channel Attacks
                [Attack{er} capability
                            [Passive, bottom color=gray!10
                            ]
                            [Active
                            ]
                ]
                [Information leaks
                            [Unintended, bottom color=gray!10
                            ]
                            [Intended, bottom color=gray!10
                            ]
                ]
                [Exploited properties
                            [Logical
                                    [Mem. {access patterns}]
                            ]
                            [Physical, bottom color=gray!10
                                    [Timing]
                                    [Sound]
                                    [Heat]
                                    [Power consumption, bottom color=gray!10]
                                    [EM , bottom color=gray!10]
                            ]
                ]
                [Required access
                            [Local]
                            [Physical, bottom color=gray!10]
                            [Remote, bottom color=gray!10]
                ]
                [Profiling phase 
                            [Non-profiled, bottom color=gray!10]
                            [Profiled, bottom color=gray!10]
                ]
        ]
    \end{forest}
\end{minipage}}
\caption{{A general taxonomy of SCA attacks}. Gray boxes highlight the features covered by the attacks {surveyed} in this paper.
\label{fig:taxonomy_SCA}}
\end{figure}
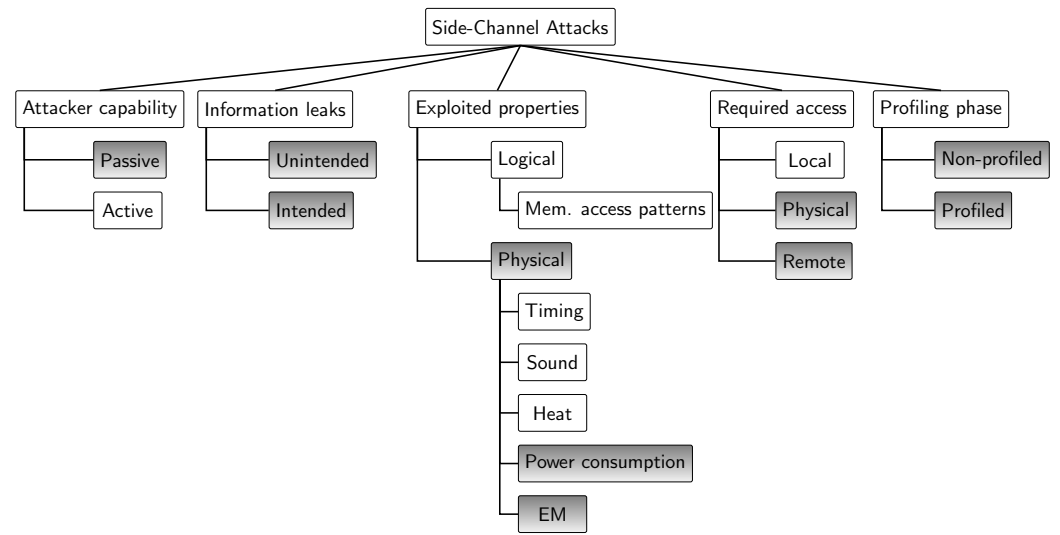

In order to help categorise SCA {attacks} against {NN} implementations {as} studied in this work, a {general} taxonomy {for SCA attacks is proposed and} summarised in \figurename~\ref{fig:taxonomy_SCA}. The features in gray correspond to the ones covered by the attacks {surveyed} in this work. This taxonomy is based on a set of different features:
\begin{itemize}[leftmargin=7.5mm,labelsep=0.5mm,topsep=3pt]
    \item \textbf{Attacker capability:} Two main attack families can be distinguished: \textit{passive} and \textit{active}. In the former one, the attacker is passive in the sens{e} that he/she can only observe the information leaked and investigate changes in these leakages in order to correlate them to possible causes. For instance, a difference in the execution time (or power consumption) is a consequence of different computation{s} or different processed data. By modeling {the} observed effects of possible specific causes, the adversary is able to learn useful information related to the actual computation on the target device. In the second family of attacks, the adversary is able to influence the target device via a side{-}channel, for instance, by modifying its environmental conditions~\cite{spreitzer2017systematic}. Note that feeding the target with arbitrary inputs, for instance, by triggering the encryption of a specif{ic} text or in our case querying a {D}NN with any input image is, even for a potential attacker, a normal and authorised action and is not considered an active attack. In this study we focus on passive SCA {attacks} on {D}NN implementations.
    
    \item \textbf{Information leaks:} Following the work {from Spreitzer et al.}~\cite{spreitzer2017systematic}, information leaks can be categorised into two types. On the one hand, \textit{traditional} leaks are the most studied and define the \textit{unintended} information leak{ag}e inherent to hardware implementation{s} on every system. These include execution and delay time, power consumption, EM emanation, heat, sound, etc. These leaks are considered unintentional as they naturally result {from} the computation on a hardware device without any intention of designers or developers on {creating} these {side-channel sources}.
    On the other hand, in addition to unintended information leak{ag}e, devices provide more features such as the ability of providing run-time{ and} \textit{intended} information of the system. For instance, information on embedded sensors (e.g., power consumption) is provided for the optimisation of the resource{s} management. This results {i}n \emph{publishing} data on purpose, which can leak information and can be exploited in the same manner {as} unintended information leaks. {The s}urveyed attacks exploit information {leaked from} unintended {side-channel sources}. 
    
    \item \textbf{Exploited properties:} Attacks exploit different kind{s} of properties for their observations. We distinguish \textit{logical} properties, such as memory {access patterns}, from \textit{physical} ones including timing, sound, heat, etc. {Specifically}, the studied attacks have exploited physical properties {such as} power consumption and EM emanation. The nature of the exploited property may or may not condition the access required to perform the attack.
    
    \item \textbf{Required access:} First, the attacker can have local access to the system in \textit{software-only attacks}, where only an adversary application co-located with the victim on the same hardware resources is required (e.g., attacks observing the access patterns/times on shared resources such as cache memories). Second, the attacker can {be in} physical proximity to the system in order to monitor power consumption or EM {emissions} (i.e., this is a hardware attack). A third category called \textit{remote attack{s}} is distinguished. This latter defines hardware attacks performed from software, for instance, attacks observing power consumption from embedded sensors information accessed by software. Note that attacker access on the target system is irrespective to the exploited property. For instance, an adversary application co-located with the victim and sharing hardware resources can exploit timing physical properties by observing timing through changes on the state of the shared cache. 
    In this work physical and remote access-based attacks are covered.
    
    \item \textbf{Profiling phase:} A profiling phase prior to the actual attack might be necessary for certain attacks. During this phase, the attacker characterises the device behaviour and side-channel information leakage in an ideal environment where the adversary has full control and knowledge of the victim. This includes the access to an exact copy of the system, full knowledge of the neural network implementation on the target device and full control of the inputs and access to the output. The adversary is therefore capable of extensively characterising the device according to the observed side-channel information in order to distinguish different \textit{profiles} (also called \textit{templates}. In the second phase{,} which corresponds to the actual attack, the adversary has limited knowledge and/or control on the victim (e.g., limited knowledge on the neural network implementation details or limited control on the inputs) and compares the observed side-channel information to the prior characterisation in order to determine the most probable profile. Attacks requiring a profiling phase are also known as \textit{template attacks}. 
    
 \end{itemize}

Finally, different SC{A} main techniques and variants exist depending on the exploited physical properties. We focus on the exploitation of power consumption and EM{ that are introduced in the following subsection}.

\subsection{SCA Techniques Covered in This Work}
When examining common analysis techniques, two main SCA categories are distinguished.
First, \textbf{Simple Power/EM Analysis} (SPA and SEMA{,} respectively) is the most straightforward SCA technique as it directly uses the coarse-grained data dependencies in power/EM emanations for deducing the secret value. It exploits a single or a few traces{ and is based on direct \emph{naked-eye} observations}.
Second, \textbf{Differential Power/EM Analysis} (DPA and DEMA) is an advanced technique {that} requires statistical analyses on many traces collected from sets of operations to deduce the secret value through fine-grain{ed} data dependencies in power/EM signatures~\cite{kocher1999differential}. By creating a hypothes{i}s on physical signatures or intermediate data, it tests for dependencies between the actual measurement traces and the hypothesis.
Secret data are considered {in} small parts (e.g., byte{-}by{-}byte in cryptographic algorithms, weight{-}by{-}weight or layer{-}by{-}layer in our NN scenario) in order to reduce the complexity of the attack.
In general, hypothes{e}s on the signatures are based on leakage models following the attacker's \textit{intuition} and knowledge. There exist models that are commonly used for representative hardware targets (e.g., Hamming Weight (HW) or Hamming Distance (HD)).
Finally, \textbf{Correlation Power/EM Analysis} (CPA/CEMA) is a special case of DPA/DEMA that was introduced {by Brier et al.}~\cite{brier2004correlation}. It applies statistical methods, such as correlation {analysis}, to distinguish the correct secret value among the hypothes{e}s.

\section{Threat Model and {Attack} Motivation} 
\label{sec:Threat_model_and_motivation}

This section presents different assumptions made on the attacker access capabilities to the target device, on the victim {D}NN model {and on the} attacker knowledge on the {D}NN and its motivation for the attack.  

\textbf{Attacker access assumptions:}
{The D}NN{'}s execution environment is shifting from cloud servers to edge devices{,} including low-power processors and FPGAs{,} and are expected to be {found} everywhere in the IoT. We focus on physical/remote access-based SCA {attacks} (refer to Section~\ref{subsec:Generalities} and \figurename~\ref{fig:taxonomy_SCA}).
In the considered threat scenario, the attacker exploits physical measurements (limit{ed} to power/EM {in this survey}) via a physical{ and close by} (or remote) access to the target device. The target device implements a pre-trained {D}NN model. The attacker passively observes and analyzes physical measurements during the inference operation. 

\textbf{Model assumptions:}
In this work, it is assumed that the manipulated data is not encrypted. 
It is important to notice that few recent works evaluate the feasibility of deep learning computing over encrypted data, manipulating encrypted data in the inference stage, training over encrypted data and/or using encrypted parameters in the models, i.e., by homomorphic encryption~\cite{qaisarahmadalbadawi2020towards,gilad2016cryptonets}. This approach addresses data confidentiality, especially in the case where a client uses distant resources to perform the inference stage~\cite{Chillotti}. This approach is also studied for training models over encrypted data in order to protect the training data set that might be confidential. However, even if these first endeavors are promising, homomorphic encryption introduces a great computation overhead that cannot be affordable in today's systems. 
In this work, manipulated data is assumed to be non encrypted. Furthermore, {the} considered target {D}NN models do not include any protection against SCA.

\textbf{Attacker knowledge assumptions:}
\textit{(i) Black-box scenario.} In the standard assumptions on the attacker knowledge, the model, its architecture and parameters and training data set are considered secret and therefore not known by the attacker. However, he/she is able to query the network with arbitrarily chosen inputs and can observe the resulting output.  
Depending on the attack, the scenario and the capabilities of the attacker might vary.
However, a common assumption is to consider the training set as confidential. This is a classical and logical assumption as, in our case, the model is pre-trained before it is deployed on the target device.
\textit{(ii) Gray-box scenario.} As for the model, according to the attack it might be partially known by the attacker. The attacker may also partially know the following: architecture layout, number of layers, neurons, etc., and/or known parameters such as weights and/or biases (\textit{gray-box {model}}). This is not the standard assumption, which would be to consider the entire model confidential and, for instance, if it is industrially provided (\textit{black-box {model}}). However, some work{s} consider that partial information about the model can be recovered through prior SCA and that this partial knowledge would facilitate further analysis aiming, for instance, at recovering private input data~\cite{wei2018know}. Finally, inference input{s} might also be considered private in some scenarios. Consider for instance the scenario where encrypted medical images are provided to the inference engine in order to prevent eavesdropping during data transfer~\cite{wei2018know}. In this case, the attack aim is to retrieve the private inference input{s}. 
\textit{(iii) White-box scenario.} Finally, in a white-box scenario, the adversary is assumed to have complete knowledge on the network model and training data set.

\textbf{Aim and motivation of the attack:}
Why {use SCA techniques to} analyze {D}NN implementations? Different attack objectives and motivation scenarios can be distinguished and, according to each of them, the threat model might vary.
\begin{enumerate}[label=(\roman*),leftmargin=7.5mm,labelsep=0.5mm,topsep=3pt]
    \item \textbf{Reverse engineering.} In the context of industrial {D}NNs, Intellectual Property (IP) models are a secret. A use-case for SCA in this context is an attacker who has a (legal) copy of a pre-trained network but does not have any details on the model architecture, parameters and, in general, training set (black-box). As the fine tuning of parameters, for instance weights, is of great importance in optimised networks accuracy and is currently one of the main challenges, commercial network details are kept confidential. Reverse engineering the network details would allow an attacker, who might be a competitor, to replicate the commercial IP resulting in substantial consequences including counterfeiting and economic losses. Two main objectives that will serve for evaluation are the following: (i) accuracy of the extracted information when compared to the {victim} network and (ii) task equivalence searching to obtain {similar} results than the victim model. Th{ese} evaluation metrics are further discussed in Section~\ref{sec:Discussion_and_future_research_leads}.   
    It has been proven that SCA enables the recovery of {the} network models architecture and parameters including activation functions, number of layers, neurons, output classes and weights. The standard assumption for this type of attacks is that an attacker has no knowledge on the network but can feed it with chosen inputs and has access to the outputs. 
    Furthermore, in the scenario of commercial models, data being used to train the model are, in general, kept confidential as well. Inde{ed}, in some cases, these data can be extremely sensitive, for instance, for models trained on medical patients data records.

    \item \textbf{Data theft.} Through SCA, inference input data (data to be classified) can be recovered by directly hampering data confidentiality and user's privacy. In privacy-preserving applications such as medical image analysis, patients' data privacy requires the utmost attention. In general, as {the} network architecture becomes more and more complex, deducing the input data by observation of outputs is not considered feasible. In the literature, SC{A}-based works assume partial or entire knowledge on the network architecture and parameters in order to be able to retrieve inference input data~\cite{wei2018know,batina2019csi}. A possible use-case is an attacker who does not know nor control the inputs/outputs but knows the network as it is publicly available or as it has been previously reverse engineered such as in~\cite{batina2018csi}. 
\end{enumerate}

\section{Physical SCA {Attacks} on {D}NN{s}}
\label{sec:Physical_SCA_on_NN}
Since {the} last couple of years, research has shown that embedded {D}NNs are vulnerable to physical SCA {attacks}. 
{The} first very recent endeavors on classifying existing SCA attacks on {D}NNs have {only} very recently been published. Considering a different perspective from this paper, these first efforts are complementary and show the exten{t} of this emerging research field and the necessity to have a broad overview of current attacks. However, to the best of our knowledge, no previous work has focused on physical power and EM-based SCA against embedded {D}NN{s}.

H. Chabanne et al.~\cite{chabanne2021side} present SCA on {C}NNs aimed at extracting the model architecture. The attack considered scenario is limited to {D}NN models on mobile phones as well as hosted by a cloud provider in the context of ML as a service (MLaaS). Contrary to this work, we do not consider MLaaS but we focus on {the} IoT context and on pre-trained models deployed on low-cost processors or FPGA{s} that are physically accessible by the attacker. {Specifically, the authors of this work}~\cite{chabanne2021side} {provide} an overview on existing timing and local access-based attacks (refer to the taxonomy proposed in Section~\ref{subsec:Generalities} and \figurename~\ref{fig:taxonomy_SCA}) in which the adversary application is co-located on the same hardware machine as the victim, thus sharing hardware resources and particularly cache {memories}. In this scenario, physical proximity to gain access to physical measurements is not required. In contrast to this work, we focus on physical access-based SCA (note that they might be performed remotely as presented further in this section). 
Therefore, {SCA based on} local access and {exploitation of} logical properties are out of the scope of this work (see \figurename~\ref{fig:taxonomy_SCA}).

Xu et al.~\cite{xu2021security} focus on active attacks that are out of the scope of our work. {The a}uthors particularly {consider} hardware Trojan insertion and fault injection. They gave an overview on attacks aimed at extracting models as well as corrupting models and outputs. Contrary to this work, we focus on passive attacks exploiting {side-channel} information.

Finally, {a} very recent white paper is available~\cite{joud2021review} {that} provides an overview of different literature endeavors on the extraction of training data and {D}NN model{s}. Reviewed work{s} included some SCA attacks, as well as purely software attacks (e.g., adversarial example-based attacks also referenced as application programming interfaced-based attacks). 
Similar{ly}, recent work~\cite{mittal2021survey} considers, among others, physical SCA in its review. 
Contrary to this paper, {the} authors did not specially focus on power and EM SCA and did not consider recent physical SCAs that are remotely performed. However{,} all these endeavors are complementary to our work. In this paper the aim is to provide an understandable yet detailed view and classification{,} focusing on power and EM SCA {attacks} that are not limit{ed} to the extraction of training data set and models. 

\subsection{Taxonomy of SCA {Attacks} Against {D}NN Implementations}

\begin{figure}[H]\widefigure
\begin{minipage}{9cm}
\hspace{-8.5cm}\begin{forest}
        family tree too,
        [Taxonomy of SCA attacks on {D}NN implementations
                [Aim of the attack
                            [Reverse engineering
                                    [NN architecture
                                        [\#/type of layers]
                                        [\# Neurons]
                                        [\# Output classes]
                                        [\# Activation]
                                    ]
                                    [Parameters
                                        [\# Weights]
                                    ]
                                    [Training data
                                    ]
                            ]
                            [Data theft
                                    [Input data]
                            ]
                ]
                [{Attacked} DNN
                            [MLP]
                            [CNN
                                [BNN]
                            ]
                ]
                [Data set
                            [Real data]
                            [Synthetic data
                                [MNIST]
                                [CIFAR-10]
                                [ImageNet]
                            ]
                ]
                [Physical target
                            [\textmu C]
                            [FPGA]
                ]
                [Physical measurement 
                            [Power
                            ]
                            [EM 
                            ]
                ]
                [Required access
                            [Physical
                            ]
                            [Remote
                            ]
                ]
                [Threat scenario
                            [White-box]
                            [Gray-box]
                            [Black-box]
                ]
        ]
    \end{forest}
\end{minipage}
\caption{Taxonomy of SCA attacks on {D}NN implementations studied in this work.
\label{fig:taxonomy_attacks}}
\end{figure}
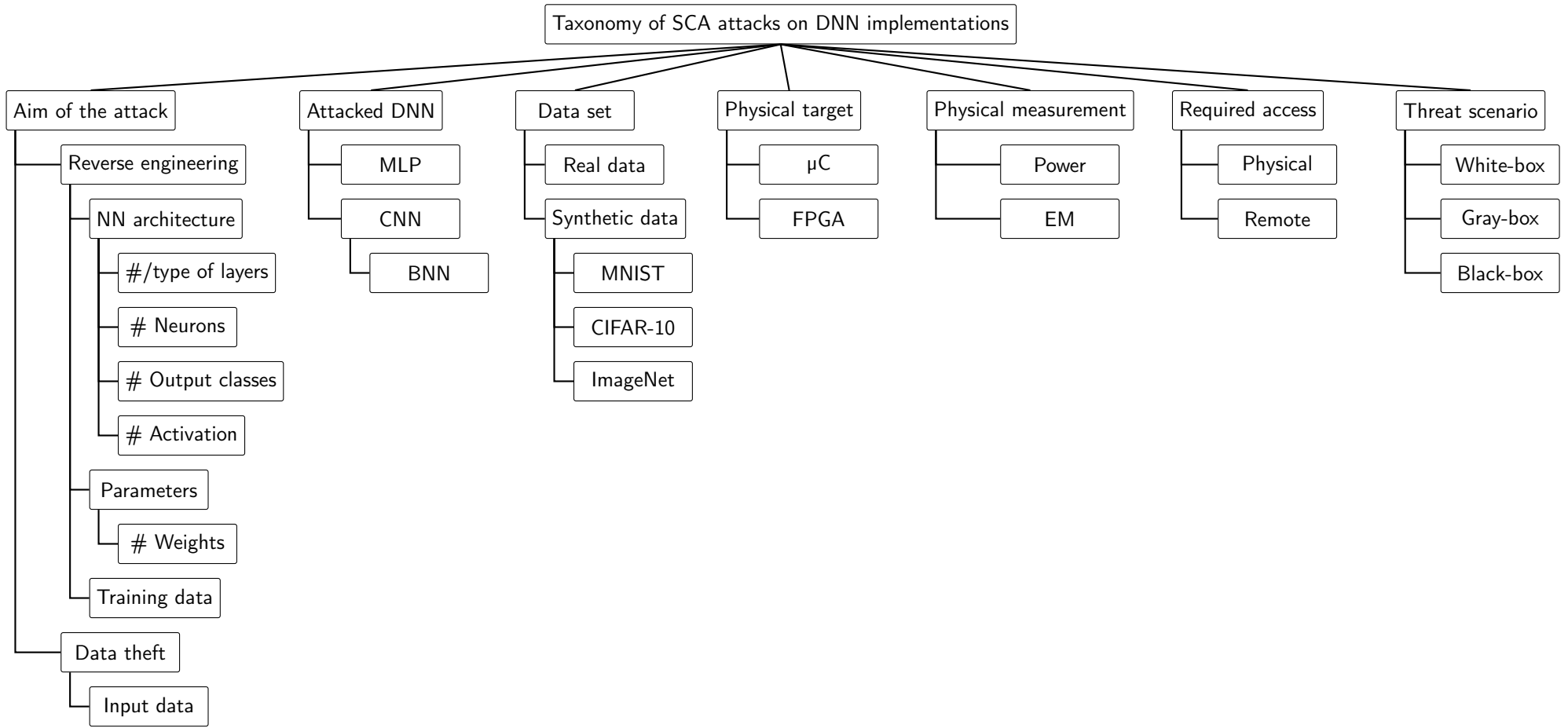

In this section{,} SCA attacks {from the literature} on {D}NN implementations within the considered threat scenarios {from} Section~\ref{sec:Threat_model_and_motivation} are studied. In order to provide a global overview of existing attacks, they are classified according to different features. 
The features on the proposed taxonomy for SCA in general, presented in Section~\ref{subsec:Generalities} and summarised in \mbox{\figurename~\ref{fig:taxonomy_SCA}}, can be considered. However, this work is limited to a subfamily of SCA according to the considered scenarios {with the} relevant features in this scenario highlighted in grey. Moreover, additional features specific to the {D}NN domain and the considered threat scenario assumptions need to be taken into consideration. First, features introduced in Section~\ref{sec:Threat_model_and_motivation} include the aim of the attack and the threat scenario. Second, features regarding the {D}NN model and physical target have to be taken into account. Indeed, by focusing on DNN, literature attacks have targeted MLP, CNN and BNN models. Moreover, for experimentation, real data or synthetic data set can be used. Not{e} that the great majority of the studied works {are} limited to simple synthetic data sets (e.g., MNIST, CIFAR-10 and ImageNet data sets). 
Finally, literature attacks have targeted micro-controllers and/or FPGA-based {D}NN implementations.
These features are gathered in \figurename~\ref{fig:taxonomy_attacks}, providing the complete taxonomy for SCA on {D}NN implementations.

\startlandscape
\finishlandscape

In the remainder of this section, we review the literature attacks and highlight information following {t}he proposed taxonomy features. 
Attacks are presented depending on features considered important for an understandable classification: first, according to the aim of the attack; second, according to the physical target (as the attack vector exploited might depend on the target specifications); finally, according to the physical measurement. We discuss {the} attack{'}s threat model{s} and limitations. In order to favor reproducible endeavors that can be perfomred by possible readers of this work, we highlight, for each attack implementation, the hardware target and network model and reference the implementation details' documentation, articles and/or {code} sources whenever they are specified. 
The proposed classification overview is provided in Table~\ref{table:Categorization_of_attacks,_NS_stands_for_non_specified.}, where each studied attack is individually considered. A complementary overview is presented in Table~\ref{Classification_of_attacks_according_to_each_feature.}, which is organised per studied feature. Additionally, it provides information on exploited vectors and metrics used for the evaluation of the attack on both fronts in terms of efficiency and cost. 

\subsection{Recovering Network Architecture and Parameters}
\subsubsection{Micro-Controller {Targets}}
\label{subsec:Recovering_architecture_model}

\textbf{EM side-channels:} Batina et al.~\cite{batina2019csi} presented an attack that was able to reverse engineer the network architecture and recovered key parameters such as the pre-trained weights and activation functions as well as the number of layers and neurons per layer through the analysis of EM signatures. The entire model architecture and training data set are considered private and only tested inputs and outputs that are known by the attacker (black-box). 
{The a}uthors targeted a pre-trained model of an MLP consisted of three layers with different number of neurons per layer and a CNN model operating on real numbers introduced {previously}~\cite{lai2018cmsis} {that} consists of three convolution layers, three pooling layers and one fully connected layer. {The a}uthors use MNIST~\cite{deng2012mnist} and CIFAR-10~\cite{cifar-10} data sets (for MLP and CNN models{,} respectively) and conducted their experiments on two micro-controllers{,} which are Atmel ATmega328P and ARM Cortex-M3 32-bit. 
First, {the} authors analysed the EM {side-channel} signatures of the most commonly used activation functions (ReLU, sigmoid, tanh and softmax) by randomly varying the input within a certain range (\{$-$2,2\}). From the EM traces, {the} authors measure the timing of the activation function and show that these functions result on very different computing times for the same inputs{, which are} easily distinguishable. This enables the straightforward deduction of the function that had been used and ReLu was the fastest.
Second, the proposed attack targets the data bus when loading, from memory, the secret pre-trained weights and exploits the characteristic of data buses in micro-controllers of pre-charging to all `0's before every instruction. This allows the observation of the new EM leakage signature after each regular load of the weights and permits modelling based on the weight's Hamming Weight (HW). {The a}uthors target the multiplication $m = input \cdot weight$, where the $input$ is known by the attacker, to deduce the secret $weight$. Following a CEMA-based method, they correlate the activity of the predicted $m$ for all hypothes{e}s of the $weight$ to distinguish the correct weight value among their hypothes{e}s. To limit the problem, {the} authors assumed that the weight will be bounded in a certain range chosen by the attacker. {The a}uthors recover the mantissa bit of the weight before recovering separately the sign and exponent bits. These latter two further narrow the candidate list and enables them to obtain an actual value of the weight. The weights were recovered with small precision errors (at third place after decimal point). 
Finally, by using SEMA on MLP (three layers, six neurons and sigmoid activation function), {the} authors were able to easily distinguish multiplication operation{s} and activation function{s} to count the number of neurons. This is, however, not possible for deeper NN{s}. {The a}uthors then proposed a CEMA-based technique to identify layer boundaries on CNN{s}. 
\startlandscape
\begin{specialtable}[H]
\widetable
\tablesize{\footnotesize}
\caption{Categorisation of attacks; \emph{NS} stands for non specified.} 
\label{table:Categorization_of_attacks,_NS_stands_for_non_specified.}
\setlength{\cellWidtha}{\columnwidth/7-2\tabcolsep-0.3in}
\setlength{\cellWidthb}{\columnwidth/7-2\tabcolsep-0.0in}
\setlength{\cellWidthc}{\columnwidth/7-2\tabcolsep-0.2in}
\setlength{\cellWidthd}{\columnwidth/7-2\tabcolsep-0.9in}
\setlength{\cellWidthe}{\columnwidth/7-2\tabcolsep-0.0in}
\setlength{\cellWidthf}{\columnwidth/7-2\tabcolsep+0.4in}
\setlength{\cellWidthg}{\columnwidth/7-2\tabcolsep+0.9in}
\scalebox{1}[1]{\begin{tabularx}{\columnwidth}{>{\PreserveBackslash\centering}m{\cellWidtha}>{\PreserveBackslash\centering}m{\cellWidthb}>{\PreserveBackslash\centering}m{\cellWidthc}>{\PreserveBackslash\centering}m{\cellWidthd}>{\PreserveBackslash\centering}m{\cellWidthe}>{\PreserveBackslash\centering}m{\cellWidthf}>{\PreserveBackslash\raggedright}m{\cellWidthg}}
\toprule
  \textbf{Attack} & \textbf{Aim of the Attack} & \textbf{Attacked Network} & \textbf{Data Set} & \textbf{Physical Target} &  \textbf{Physical Measurement, Technique} & \multicolumn{1}{c}{\textbf{Requirements/Limitations}} \\
  \midrule
  Batina et al.'19~\cite{batina2019csi} & Recover network architecture and parameters (activation function, number of layers and neurons and weights) & MLP, CNN & MNIST & Atmel ATmega328P, ARM Cortex-M3 & EM (SEMA, CEMA) & Minimal (black-box)\\ \midrule
  
  Maji et al.'21~\cite{maji2021leaky} & Recover model weights, biases & CNN, BNN & MNIST & Atmel ATmega32P, ARM Cortex-M0+, custom-designed RISC-V & SPA (timing) & Knowing the network architecture (gray-box) and disabling all peripherals and methodology specific to \textmu C\\ \midrule 
  
  Yoshida et al.'19~\cite{yoshida2019model} & Recover model weights & MLP & \emph{NS} & FPGA (\emph{NS}) & EM (CEMA) & Intention paper \\\midrule
  
  Yoshida et al.'20~\cite{yoshida2020model}, Yoshida et al.'21~\cite{yoshida2021model} & Recover model weights & Systolic array & {\emph{NS}} & Xilinx Spartan3-A & Power (CPA, chain-CPA) & Knowing the network architecture (gray-box) and the accelerator architecture; only systolic array is implemented\\ \midrule
  
  Dubey et al.'20~\cite{dubey2020maskednet} & Recover model weights & BNN (adder tree) & MNIST & Xilinx Kintex-7 & Power (DPA) & Knowing the network architecture (gray-box) and hardware implementation details \\ \midrule
  
  Yu et al.'20~\cite{yu2020deepem} & Recover network architecture and weights & BNN & CIFAR-10 & Xilinx ZynqXC7000 & SEMA, adversarial training & Black-box, restriction of certain parameters to few values and identical hidden layers\\\midrule
 Xiang et al.'20~\cite{xiang2020open} & Distinguish among different NN models and parameters sparsity & CNN & ImageNet & Raspberry Pi & Power, SVM classifier & Knowing the set of possible network architectures, using known pruning techniques and using non fine-tuned models once trained (gray-box)\\ \midrule
  Wei et al.'18~\cite{wei2018know} & Recover network inputs & BNN & MNIST & Xilinx Spartan-6 LX75 & Power ({template attack}) & Specific to line buffer, suitable for plain background images and knowing the network architecture and parameters (white-box)\\\midrule
  
  Batina et al.'18~\cite{batina2018csi}, Batina et al.'19~\cite{batina2019poster} & Recover network inputs & MLP & MNIST & ARM Cortex-M3 & EM (HPA, DPA) & Knowing the network architecture and parameters (white-box)    \\\midrule
  
  Maji et al.'21~\cite{maji2021leaky} & Recover network inputs & CNN (zero-skiping, normalised NN), BNN & MNIST & Atmel ATmega328P & Power (SPA) & Knowing the model architecture and parameters (white-box), disabling all peripherals and methodology specific to \textmu C \\\midrule
  
  Moini et al.'21~\cite{moini2020remote}, Moini et al.'21~\cite{moini2021power} & Recover network inputs (remote attack) & BNN & MNIST & Xilinx ZCU104, VCU118 & Power (remote) & Knowing the network architecture and parameters (white-box) and the adjacent location to victim module is required\\
\bottomrule
\end{tabularx}}
\end{specialtable}
\finishlandscape

\begin{specialtable}[H]
\tablesize{\footnotesize}
\caption{Classification of attacks according to each feature.}
\label{Classification_of_attacks_according_to_each_feature.}\label{tabref:beverages-873261-t002}
\setlength{\cellWidtha}{\columnwidth/2-2\tabcolsep-1.2in}
\setlength{\cellWidthb}{\columnwidth/2-2\tabcolsep+1.2in}
\scalebox{1}[1]{\begin{tabularx}{\columnwidth}{>{\PreserveBackslash\centering}m{\cellWidtha}>{\PreserveBackslash\raggedright}m{\cellWidthb}}
\toprule
  \textbf{Aim of the Attack} & \\ \midrule 
  Reverse engineering & Recovering model layout param~\cite{batina2019csi,yu2020deepem}, weights~\cite{maji2021leaky,batina2019csi,yoshida2019model,yoshida2020model,yoshida2021model,yu2020deepem,dubey2020maskednet}, biases~\cite{batina2019csi,maji2021leaky} and distinguishing among model architectures and parameters sparsity ratios~\cite{xiang2020open}\\
  Data theft & Recovering inputs~\cite{wei2018know,batina2018csi,batina2019poster,maji2021leaky,moini2020remote,moini2021power}\\\midrule
  \textbf{Attacked NN} &\\ \midrule
  MLP &~\cite{batina2018csi,batina2019csi,yoshida2019model,batina2019poster}\\\midrule
  CNN &~\cite{batina2019csi,maji2021leaky}, Zero-skiping, normalised~\cite{maji2021leaky}, Systolic array only~\cite{yoshida2020model,yoshida2021model}, AlexNet, InceptionV3, ResNet50, ResNet101~\cite{xiang2020open},  \mbox{BNN~\cite{dubey2020maskednet,yu2020deepem,wei2018know,moini2020remote,moini2021power,maji2021leaky}}, ConvNet, VGGNet~\cite{yu2020deepem}\\ \midrule
  \textbf{Dataset} & \\ \midrule
  MNIST &~\cite{wei2018know,batina2018csi,batina2019poster,batina2019csi,dubey2020maskednet,moini2020remote,moini2021power,maji2021leaky} \\ \midrule
  CIFAR-10 &~\cite{batina2019csi,yu2020deepem,maji2021leaky} \\ \midrule
  ImageNet &~\cite{maji2021leaky,xiang2020open} \\ \midrule
  \textbf{Physical Target} & \\ \midrule
  FPGA &~\cite{yoshida2019model}, Spartan3-A~\cite{yoshida2020model,yoshida2021model}, Spartan-6 LX75~\cite{wei2018know}, ZynqXC7000~\cite{yu2020deepem}, Kintex7~\cite{dubey2020maskednet}, ZCU104 and VCU118~\cite{moini2021power,moini2020remote}\\ \midrule
  \textmu C & Rasberri Pi~\cite{xiang2020open}, ARM Cortex-M0+, custom-designed RISC-V~\cite{maji2021leaky}, ATmega328P~\cite{batina2019csi,maji2021leaky} and ARM Cortex-M3~\cite{batina2018csi,batina2019poster,batina2019csi}\\ \midrule
  \textbf{Physical Measurement} & \\ \midrule
  Power & SPA~\cite{maji2021leaky}, DPA~\cite{batina2018csi,dubey2020maskednet}, CPA/chain-CPA~\cite{yoshida2020model,yoshida2021model}, {template attack}~\cite{wei2018know}, power, SVM~\cite{xiang2020open} and remotely obtained~\cite{moini2021power}~\cite{moini2020remote}\\ \midrule
  EM & HPA~\cite{batina2018csi,batina2019poster}, SEMA~\cite{maji2021leaky,batina2019csi}, SEMA \& adversarial~\cite{yu2020deepem}, DEMA~\cite{batina2018csi,batina2019poster} and CEMA~\cite{batina2019csi,yoshida2019model,yu2020deepem} \\ \midrule
  \textbf{Exploits} & \\ \midrule
  Architecture design & Systolic array~\cite{yoshida2020model,yoshida2021model}, adder tree~\cite{dubey2020maskednet} and line buffer~\cite{wei2018know,moini2020remote,moini2021power} \\ \midrule
  Hardware target specificity & All-'0's pre-charge of data bus~\cite{batina2019csi} and timing extraction of individual operations~\cite{batina2019csi,maji2021leaky} \\ \midrule
  Correlation between & Parameters sparsity and power consumption~\cite{xiang2020open}, latency and number of parameters/operations~\cite{yu2020deepem}, power/EM signatures and secret data processed (summations of products)~\cite{wei2018know,batina2018csi,batina2019poster,yu2020deepem,dubey2020maskednet,yoshida2020model,yoshida2021model} and activation functions~\cite{batina2019csi,yu2020deepem}\\ \midrule
  \textbf{Evaluation Metric} & \\ \midrule
  Attack accuracy & Recovered pixel-level accuracy~\cite{wei2018know,batina2018csi,batina2019poster,moini2020remote,moini2021power}, mean structural similarity index between original and recovered image (MSSIM~\cite{wang2004image})~\cite{moini2020remote,moini2021power}, input precision recognition of the network (original vs recovered)~\cite{wei2018know,batina2019csi,yu2020deepem}, average accuracy of recovered NN parameters~\cite{batina2019csi,xiang2020open,maji2021leaky} and normalised cross-correlation~\cite{moini2020remote,moini2021power}\\ \midrule
  Attack efficiency & Portion of the correct recovered values (weights)~\cite{yoshida2019model,yoshida2020model,yoshida2021model}\\ \midrule
  Attack complexity & Number of measurements required~\cite{yoshida2020model,yoshida2021model,dubey2020maskednet,moini2020remote,moini2021power}, image reconstruction complexity and memory complexity~\cite{wei2018know}\\ \midrule
  \textbf{Attacker Capabilities} & \\ \midrule
  Attacker knowledge & Network architecture~\cite{maji2021leaky,yoshida2020model,yoshida2021model,dubey2020maskednet,batina2018csi,batina2019poster}, set of possible network architectures~\cite{xiang2020open}, network parameters~\cite{wei2018know,batina2018csi,batina2019poster,xiang2020open,moini2020remote,moini2021power,maji2021leaky}, used pruning techniques~\cite{xiang2020open} and hardware implementation details~\cite{dubey2020maskednet,yoshida2020model,yoshida2021model} \\ \midrule
  Control on inputs &~\cite{batina2019csi,yoshida2019model,yoshida2020model,dubey2020maskednet,yoshida2021model,xiang2020open,yu2020deepem,maji2021leaky} \\ \midrule
  Disabling all peripherals &~\cite{maji2021leaky} \\ \midrule
  \textbf{Assumptions Made to Facilitate the Attack} & Limited set of possible network models~\cite{xiang2020open}, limited set of possible filter-sizes~\cite{yu2020deepem}, recovering reduced precision values~\cite{batina2018csi,batina2019csi,batina2019poster} and input images with easily distinguishable background from foreground~\cite{wei2018know,moini2020remote,moini2021power} \\\bottomrule
\end{tabularx}}
\end{specialtable}

The entire methodology was repeated layer by layer and weight by weight to recover the entire model architecture. For validation, {the} authors compare the accuracy of the original network and the reverse engineered one. Results show comparable accuracy with an averaged loss of precision of 0.01\% and average weight error of 0.0025\% for the MLP model and 0.36\% loss of precision for the deeper CNN{s}. 
\textbf{Discussion and limitations:}
As highlighted by the authors, the assumption{s} on this attack adversary are minimal{,} which renders this attack even more serious. On the other hand, attack complexity scaling and loss of precision of the recovered model remain to be evaluated {for} larger and deeper networks. Finally, this attack exploits a {bus} pre-charge to all '0's which is a characteristic specific to micro-controllers. 

\textbf{Power:}
Maji et al.~\cite{maji2021leaky} follow the {previous} work of~\cite{batina2019csi}, adapting the methodology to power signatures and evaluating it in {D}NN models with different precision (floating point, fixed point and binary). 
By using SPA and timing techniques, {the} authors focused on recovering the \textit{micro-parameters} (called parameters in \figurename~\ref{fig:taxonomy_SCA}), specifically the weights and biases of the model, by assuming that the \textit{macro-parameters} of the pre-trained model (i.e., model layout param{eters} in \figurename~\ref{fig:taxonomy_SCA}) such as the layout of the {D}NN, number of layers and numbers of neurons per layer, and activation function are already known by the attacker (gray-box scenario). The attacker is also able to query the network with crafted inputs. 
Similar {to}~\cite{batina2019csi}, this work uses power signatures (instead of EM) to extract timing information in order to recover the sequence of execution of the main {D}NN operations (multiplication, activation and addition). {The a}uthors consider 8-bit integers inputs and, similar to previous work, they focus on the multiplication of known inputs with secret weights following the observation that the multiplication operation timing depends on the mantissa of the operands. A mapping of all possible input activation mantissas and the timings of the corresponding multiplication operation are built. For their evaluation, {the} authors assumed a limited range for possible input mantissas. After the mantissa bit is recovered, the sign and exponent bits are extracted. Timing analysis is performed in order to distinguish the used activation function. In order to extract weight values, {the} authors use the concept of zero-crossover input~\cite{lowd2005adversarial} of input/weight pair. However{,} zero-crossover inputs are obtained only when bias{es} and weight{s} are of opposite signs. Therefore, many weights and biases cannot be uniquely determine{d}. In these cases the attacker would need to determine a different variant of zero-crossover inputs under different conditions. The proposed methodology was adapted for CNN{s} of different precision. The results relative to two layer CNN{s} using MNIST data set showed the recovery of weights and biases with <1\% error for floating point precision and exact recovery for fixed point and binary precision. 
\textbf{Discussion and limitations:} One possible limitation of this work is the asuumption that the timing of individual {D}NN operations can be extracted. This assumption might be true in micro-controllers but not in hardware accelerators when operations are executed in parallel. Moreover, in order to render the power consumption observations easier during the inference process, {the} authors were required to disabled peripherals (i.e., interrupt controllers, serial communication interface, data converters). Finally, this attack assumes prior knowledge of the {D}NN architecture. However, other attacks in the literature have already succeeded in recovering these coarse-grain features and {the} authors argue that these could be used prior to this work. On the other hand, this works stands out by the various setup configurations considered for the evaluation of their extraction method: different target micro-processor platforms (ATmega328P, ARM Cortex-M0+ and {a} custom-designed RISC-V chip~\cite{zhao2017accelerating}){;} {and D}NNs {with} different precision (floating point, fixed point and binary) using MNIST data set.

\begin{figure}[H]
\resizebox{.9\linewidth}{!}{\includegraphics{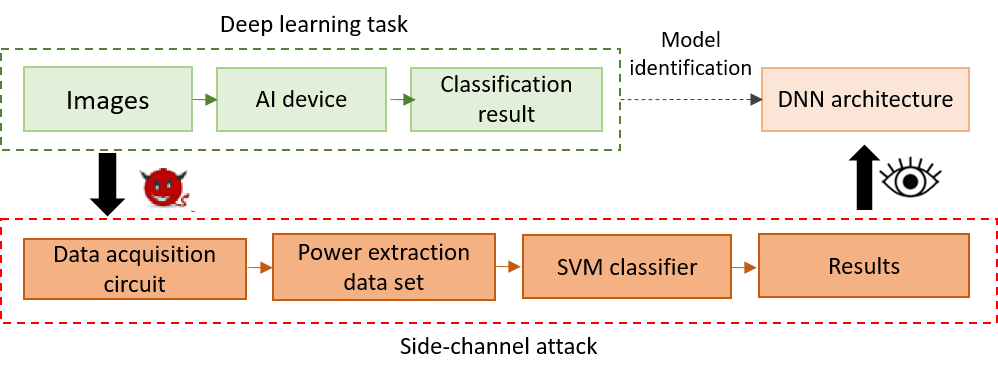}}
\caption{Framework from Xiang et al.~\cite{xiang2020open} {(artwork reworked)}.}
\label{fig:xiang2020}
\end{figure}

An emergent attack area investigates ML-based techniques against {D}NN embedded implementations using SCA. Xiang et al. propose~\cite{xiang2020open} an SCA methodology based on {a} SVM (Support Vector Machine) machine learning technique in order to identify the {D}NN model used for inference and its parameters' sparsity ratio. This work is based on the assumptions that, first, embedded {D}NNs employ, in general, existing and well known architectures so that the adversary can distinguish the used one among them{;} and, second, that as resources are limited in embedded systems, pruning techniques are often used which cause sparse parameters (zero value parameters). The intuition is then that the parameter sparsity's impact on power consumption is observable and can be analysed. Consider two different {D}NN models with identical architecture but different parameter sparsity{;} they result in very different power consumption. {The a}uthors started by (1) proposing power computational models for the main components of the DNN including convolution, pulling and fully connected layers and different activation functions based on the number of addition and multiplication operations performed by each component. Then, by ignoring other operations that would usually have a negligible effect in comparison, {the} authors built an overall power consumption model. Next, (2) assuming that pre-training and pruning techniques are usually known to adversaries, {the} authors propose a characterisation of several DNN{s} pruned to different ratios in order to obtain various models with different parameters sparsity ratios. These models are implemented in a Rasperry Pi target and the voltage and current traces of the inference operation are collected and gathered into a \textit{power-feature data set}. Part of this data set is used to train {an}other part to feed an SVM classifier that distinguishes the used DNN architecture and the sparsity ratio. The attack framework is illustrated in \figurename~\ref{fig:xiang2020}.
For evaluation {the} authors considered AlexNet, InceptionV3, ResNet50 and ResNet101 models with 0.6, 0.8 and 1.0 sparsity ratios using ImageNet data base images reshaped into 224 $\times$ 224 {pixels}. This technique achieved 96.5\% average classification accuracy for distinguishing the {D}NN model.
By extending this technique using sparsity as a feature, {the} authors are able to distinguish zero value-weights and biases without deducing their exact value and achieved 75.8\% recognition accuracy when considering the AlexNet model.
\textbf{Discussion and limitations:}
This attacks assumes that the overall architecture of the possible {D}NN used by the victim is known by the attacker as well as the pre-training and pruning techniques. It exploits pruning techniques that result in parameter sparsity. Contrary to other attacks, this work aims at distinguishing the used {D}NN  model among a set of publicly available and widely used ones without deducing the exact parameters. As a possible drawback, we can note that, afterwards, training models can be fine-tuned for optimisation. These modifications in the model could result in misclassifications of {the} SVM-based model classifier.

\subsubsection{FPGA {Targets}} 
\label{sub:FPGA}
\textbf{EM:} Yoshida et al.~\cite{yoshida2019model} introduced preliminary results of a CEMA-based attack aimed at recovering MLP model weights.
The{y target} a three layer MLP implemented in an FPGA accelerator (no specific reference is provided). Twenty 8-bit fixed point weights were considered. In th{is} scenario, weights might be encrypted when stored in an off-chip memory but would be decrypted before being processed within the {D}NN accelerator. Moreover, the adversary can freely query the network with arbitrary inputs.
{The a}uthors recover{ed} 19 out of the 20 considered weights by using a CEMA-based technique and exploited the correlation between EM observations and the HD of an accumulator register. 
\textbf{Discussion and limitations:} As {the} authors highlighted that this attack will be efficient on DNN models {for which} parameters have been encrypted, they {actually} refer to the fact that, in the chosen scenario{,} if parameters are encrypted then the adversaries can not deduce the model weights by analysing the {data transfers with} the off-chip memory. Within the accelerator, weights are non{-}encrypted when processed. While this one page paper was not able to present detailed information in order to fully understand nor reproduce the attack methodology (inputs, hardware target, required attacker knowledge on the model architecture layout and other parameters, {etc.}), it showed the first endeavor for SCA {attacks} on FPGA-based {D}NN implementations aimed at recovering model weights. 

\begin{figure}[H]
\resizebox{\linewidth}{!}{\includegraphics{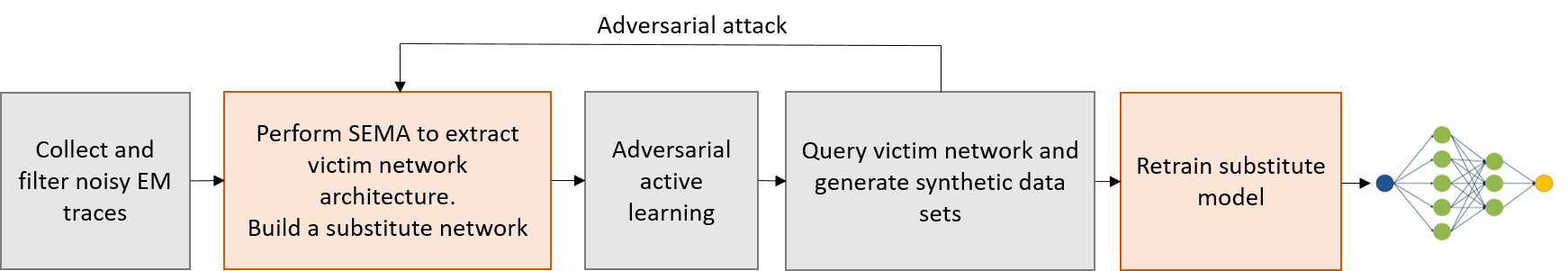}}
\caption{Overview of the attack proposed by Yu et al.~\cite{yu2020deepem} {(artwork reworked)}.}
\label{fig:yu2020deepem}
\end{figure}

Yu et al.~\cite{yu2020deepem} show an original approach that exploited both EM {side-channels} {and} margin-based adversarial active learning methods~\cite{yu2020cloudleak} in a black-box scenario in order to recover the model architecture and parameters. In the considered scenario{: (1)} the BNN model architecture and parameters, as well as the training data set, are kept confidential{; and (2)} the adversary can query the network with arbitrary inputs that might be adversarial, observe inference results{,} measure EM signatures during inference operations and observe the results. In the first part of this work, {the} authors aimed at recovering the topology of the network architecture including the number and depths of convolution, pooling and fully-connected layers through EM side-channel information by applying a SEMA-based technique. {The a}uthor{'}s intuition is that {the} EM field that is generated is proportional to the transition rate of each layer{,} which is related to {the} computations performed in each layer. Since the parameters determine the execution time of the computation of each layer, the temporal behavior of each layer and, therefore, the transition are proportional to its parameters. {The a}uthors modelled the approximation number of parameters for convolution and fully-connected layers according {to} the performed operations and parameters{. The a}uthors show that, based on their analysis, it is possible to easily distinguish the depth and type of individual layers in the EM trace. In the second part of this attack, once the architecture is recovered, {the} authors use adversarial learning-based methodologies~\cite{yu2020cloudleak} in order to induce the model to output incorrect classification results. These \textit{adversarial examples} (inputs and outputs pairs, i.e., labels or confidence scores) allow the attacker to gain insightful information on the decision boundaries and are used to build synthetic data sets to train the substitute model. {The a}uthors guess the most probable architecture by comparing {the accuracy of both the substitute model and the victim network when considering} the same data set. \figurename~\ref{fig:yu2020deepem} shows {a}n overview of the proposed attack.
For their experiments, {the} authors implement a 12-layer ConvNet and a 23-layer VGGNet as well as smaller scale LeNet and AlexNet on {a} Zynq XC7000 System-on-Chip on the Pynq-Z{1} board. {A c}omparison between the accuracy of the victim and of the substitute network achieving the best accuracy results show a{n} accuracy {loss} between ~26.2{\%} and 1.1\% compared to the original victim network. 
\textbf{Discussion and limitations:} {The a}uthors highlight the difficulties of recovering architecture{s} and parameters in large scale networks. In order to {simplify} their approach and to narrow down the possible parameters' values, {the} authors made some assumptions including the limitation of the possible convolution and pooling filter dimensions to a few values and identical hidden layers.

\textbf{Power:} Yoshida et al.~\cite{yoshida2020model,yoshida2021model} (extended version) targeted a DNN accelerator composed of a systolic array circuit implemented on an FPGA. A systolic array {SA} is a {hardware architecture for parallel processing composed of a set of regular interconnected Processing Elements (PE) that work in a pipelined fashion and where different operations can be mapped. A typical application of SAs is to accelerate matrix operations}. As NN algorithms regularly require matrix multiplications, {SAs} have proven to be very efficient (high performance and low power) for FPGA-based implementations and have been widely adopted both in {academia} and the industry, including {the} Google TPU~\cite{chen2016eyeriss,jouppi2017datacenter}. 
{The a}uthors considered, for their experiments, a TPU {architecture}~\cite{jouppi2017datacenter} as well as a wavefront array architecture~\cite{kung1982systolic}{. Both} consist of 3 $\times$ 3 PEs and each performed a multiply-accumulate operation with 8-bit integers as network inputs. For this attack, a gray-box scenario is considered where the adversary is assumed to be able to query the network with arbitrary inputs {and} assumed to know the accelerator and the DNN model architecture (number of layers, nodes and activation function). Only the training data set and model weights are considered secret. The aim of the attack is to recover the pre-trained model's weights. Using a CPA-based technique, {the} authors propose an attack methodology called \textit{chain-CPA}, which consists of several sequential CPA{s} aimed at reducing noise and induces inaccurac{ies} in the observations. By using the HD leakage model{, the} authors apply {the} chain-CPA to the register storing the multiply-accumulate operation result for each PE in the array and observed the correlation between power consumption signatures and the intermediate operations' results. Their experiments were performed on the ZUIHO platform and targeted {a} Xilinx Spartan3-A FPGA. {The r}esults show that, in general{,} chain-CPA performs better than classical CPA when evaluating the correctness of the recovered weights.   
\textbf{Discussion and limitations:} This attack technique specifically targets the {SA,} which is an implementation choice. For evaluation{,} only the {SA} (wavefront and TPU) has been implemented and 100,000 power traces were required in order to recover nine integer weights. The scalability of the approach for larger and different {SAs} requires investigation. Finally, the evaluation method consist{s} of validating the correctness of the recovered weight values{,} which will not be possible in practise.

Dubey et al.~\cite{dubey2020maskednet} assume that the details of the {D}NN algorithm and hardware implementation (data flow, parallelisation and pip{e}lining) are publicly available and are therefore considered known by an attacker. {This is a reasonable assumption in many cases.} Moreover, the attacker can query the networks with arbitrary inputs. The secrecy of the system is then based on the pre-trained model weights.
In order to recover the weights, {the} authors used a DPA-based technique focusing on the switching activity of registers as their impact in power consumption is significantly greater than combinational logic on FPGA. {The a}uthors use a 10{-}stage fully-piplined adder tree for which its registers store the intermediate summation of the product of the known input pixels and the secret weights. As the number of possible weights stored in the registers increases exponentially at each stage of the adder tree, {the} authors focus on attacking the registers of the first and second stages. {The a}uthors propose a model based on {the} HD of previous and current cycles summations in each register of the second stage of the adder. They developed a cycle-accurate simulator of the adder pip{e}line in order to {simplify} their attack{,} which functions as follows{:} (i) computes per cycle the value in each register for all possible couples of weight and a certain known input{;} (ii) computes the HD of each register for each cycle{;} and (iii) computes the total power dissipation for each cycle by adding {the} HDs of all the registers. Finally, by using Pearson's correlation between the weight guess and power measurements{, the} authors show that the correct weight value can be deduced. {The a}uthors note that this method can also be applied in order to deduce the bias as well by attacking the 10th stage of the adder tree since it is added to the final summation or by attacking the activation output. This attack methodology is evaluated through the implementation of a BNN with three fully connected layers of 1024 neurons each used for MNIST digit recognition (28-by-28 pixel images) on a SAKURA-X board {with a} Xilinx Kintex-7 FPGA. The results show that 45K measurements were necessary to obtain {the} correlation between the correct weight value guess and power measurements with 99.99\% confidence.
\textbf{Discussion and limitations:} {In addition to the attack strategy, t}he main contribution of this paper is {also} the proposal of a masking-based mitigation against power-based SCA (discussed in Section~\ref{sec:Current_countermeasures}). The implementation of this attack is also used for evaluation purposes. This attack requires the attacker to use a publicly available network architecture and to know the hardware implementation details, which are not often public information.


\subsection{Recovering Input Data}
\label{subsec:Recovering_input_data}
\subsubsection{Micro-Controller {Targets}}
\textbf{EM:}
Batina et al.~\cite{batina2018csi,batina2019poster} target a three layer-MLP pre-trained model by using floating-point numbers for inputs by normalising the MNIST database between 0 and 1 and conducted their experiments on {an }ARM Cortex-M3 32-bit micro-controller. {The a}uthors assume the network architecture is known by the attacker, including weights, activation function and number of layers and neurons. They target the multiplication operation in the first hidden layer where several known weights $w\_{i}$ are multiplied by each unknown single input $x$. {The a}uthors note that as the multiplication can be processed by {an }ALU o{r} by dedicated floating-point units, a solution in order to generalise their approach is to target the multiplication result update in the registers or memory. However, as $x$ will change between two measurements, it is not possible to leverage the information learned from one measurement with {subsequent} measurements. Therefore, {the} authors propose to use a single EM trace-based technique, called \textit{Horizontal Power analysis (HPA)}~\cite{clavier2010horizontal}, but using EM. The weights in the first layer are multiplied one by one with the same input $x$ and the single trace is split into $i$ sub-traces for $i$ weights in order to isolate them and to treat them separately. Then, the value of the input is statistically deduced through a DPA standard technique on the sub-traces applying Pearson's correlation. \textbf{Discussion and limitations:} In order to be reliable, this attack technique requires a large number of multiplications to recover inputs. Therefore, it cannot be applied on small networks consisting {of} less than 40 neurons in the first hidden layer according to {their experiments}~\cite{batina2018csi}. {Accordingly, {the} authors repeated the experiments for a} network architecture {that} encompassed 500 neurons in the first layer~\cite{batina2019poster}. When evaluating the accuracy of this attack, {the} authors consider it a success when the recovered input value matches two decimal places.

\subsubsection{FPGA {Targets}}
\begin{figure}[H]
\resizebox{.9\linewidth}{!}{\includegraphics{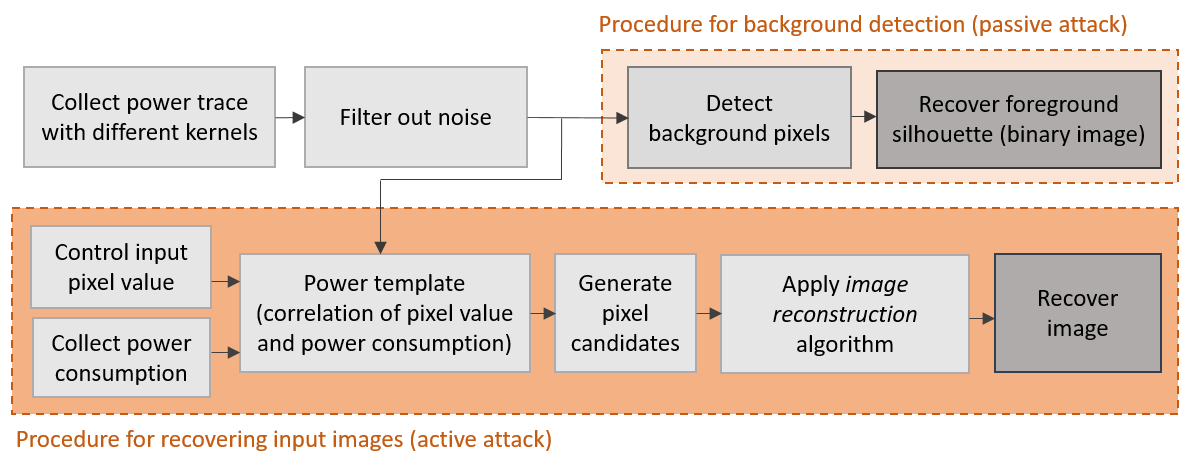}}
\caption{Overview of the attack proposed by Wei et al.~\cite{wei2018know} {(artwork reworked)}.}
\label{fig:wei2018know}
\end{figure}

\textbf{Power:}
The first attack on FPGA-based CNN{s} has most certainly been published in 2018 by Wei at al~\cite{wei2018know}. The aim of this attack was to recover private inference input data endangering data confidentiality. Based on the observation that in image related tasks, existing {C}NN architectures are often adopted and it is assumed that the adversary knows the structure of the neural network (i.e., the architecture layout, number of layers and configuration of each layer, filter dimension, number of input and output feature maps in the first convolution layer) and the input image size. Wei et al. considered two different scenarios. First, a \textit{passive} attack in which the adversary can only observe power traces during the inference of the victim input and, second, an \textit{active} attack where the adversary has the extra ability to correlate the power consumption to the input pixels by querying the inference operations with their arbitrary inputs (both scenarios' procedures are illustrated in \figurename~\ref{fig:wei2018know}).

This attack targets a line buffer{,} which is a specific hardware component {that stores image rows to} execute the convolution operation in the first layer. 
The first convolution layer is targeted as it directly processes the raw input data and {the} collected power is closely related to it. Line buffers consist of a computing engine and buffer lines. The buffer lines temporarily store pixels in recent lines while the computing engine computes the convolution of pixels and some filter values. By focusing on the power consumed {by} the convolution unit (which is the most consuming element), {the} author{'}s insight is that if the data inside remain unchanged between cycles, the induced internal transitions will be limited and therefore power consumption would be small. 
(1) \textit{Background detection:} The main idea of passive-based attack is then to observe the power consumption in each cycle and to determine {whether} pixels share similar values. If the data in the convolution engine remains {the} same (or similar) across cycles, the switching activity in the circuit will be low and will entail lower power consumption. With the intuition that {pixels with similar values} most probably belong to the background of the image (black pixels in considered MNIST data-set), {the} authors define a power consumption threshold in order to determine, for each cycle, if it processes background pixels. Pixel by pixel{, the} background is distinguished from the foreground, which permits the reconstruction of the silhouette of the foreground of the input image. 
(2) \textit{Recovering image details:} Instead of being limited to the detection of background pixels, an active attacker tries to recover the actual value of each pixel. The intuition is that, in {the} convolution layer, the same inputs are processed with different kernels. For a certain pixel region, the value of the pixels can be inferred from the power consumption of different kernels. In a {profiling stage}, prior to the actual attack, the attacker will feed the inference operation with arbitrary inputs in order to buil{d} a power template storing the mapping between pixel values and power consumption at each cycle when convolved with different kernels. The template will then be used to predict the value of each pixel during the actual attack stage by using an \textit{image reconstruction algorithm} based on a greedy heuristic. Using this template-based technique, {the} authors were able to retrieve the input images pixel by pixel. 
For experimentation, {the} authors implemented a{n existing} four-layer BNN accelerator~\cite{zhao2017accelerating} in a Xilinx Spartan-6 LX75 FPGA chip on the SAKURA-G board. The authors considered the MNIST database and introduced two metrics for {the} evaluation of their methodology: pixel-level accuracy and recognition accuracy. The first one evaluates if the value of each pixel of the reconstructed image is correct. In the passive attack, background pixels (black pixels) are regarded with value 0, while, in the active attack, this metric wants to compare the difference between the value of the retrieved pixel compared to the correct value in the original input image. The second metric evaluates the \textit{cognitive quality} of the reconstructed image by using them to feed a high accuracy MNIST classification model and by comparing the prediction result with its correct label. {The a}uthors evaluated two different kernel sizes in the first layer, 3 $\times$ 3 and 5 $\times$ 5. The passive attack highlighted the important effect of the kernel size and the selected threshold on the accuracy achieved by their method. This attack achieved better results for the smaller sized kernel (3 $\times$ 3): pixel-level accuracy of 86.2\% compared to 75.6\% for 5 $\times$ 5 kernel size and average recognition accuracy of 81.6\% compared to 64.5\% for 5 $\times$ 5 kernel size. The reason is that the proposed algorithm is based on {a} threshold and assumes that all the pixels within the convolution unit are identical. The presence of foreground pixels in the convolution unit results in a misclassification of background pixels that remain to be treated. These effects increase with the size of the kernel and are more or less significant according to the digit that is being processed. {The a}ctive attack achieved better results than the passive attack for both kernel sizes and showed better recognition accuracy for the smaller kernel size: 89\% for 3 $\times$ 3 and 79\% for 5 $\times$ 5 kernel size. 
\textbf{Discussion and limitations:} {the} MNIST data set is particularly suitable for this methodology as it facilitates the distinction between background and foreground pixels in plain black background MNIST images in order to reconstruct the input images. This is certainly one of the limitations of this work. The efficiency of the power model is then to be considered with messy background images, other than black and white, as we can expect in real world data sets. 
Furthermore, this attack exploits the line buffer in a convolution layer of the BNN implementation which is a specific design choice. However, the authors highlight that line buffer{s are} often adopted for optimised hardware implementations in image processing applications. Finally, the authors estimated the complexity of their passive attack as proportional to the number of cycles to compute the convolution{,} which depends on the image size. For the active attack, the complexity is estimated according to the time and memory complexity for building the power template and the complexity of the proposed image reconstruction algorithm.

\textbf{Remote power:}
Recent work has emerged using power estimations remotely obtained. Different from previously reviewed attacks, Moini et al.~\cite{moini2020remote,moini2021power} {(extended version)} considered multi-tenant FPGA {setup}s. In this scenario, several independent users remotely use an FPGA and, therefore, share its physical resources. A victim user runs a BNN inference module on the same FPGA where the adversary is running their modules, allowing the deduction of information on the co-located victim module. In these papers, the authors perform an attack similar to the passive attack {from Wei et al.}~\cite{wei2018know} and follow its methodology and assumptions on the attacker knowledge of the BNN model (knowledge on the model architecture and parameters, but the input and output are unknown and cannot be controlle{d} by the attacker). The main difference is in the fact that the authors use the observation on voltage fluctuation estimates that are remotely obtained through {custom built} voltage on-chip sensors based on precise time-to-digital converter (TDC) circuits. The usage of circuit delay in these sensors in obtaining estimates of the changes in the supply voltage in order to perform SCA has been previously studied in {a more traditional cryptographic context}~\cite{schellenberg2018inside}. Following this approach, the authors target the line buffer-based implementation of {the} convolution, which is {similar to Wei et al.}~\cite{wei2018know}, and focus on the first layer as its input is the raw data. {The a}uthors observed that the operation on foreground pixels results in significant switching activity, specifically in the adder tree of the convolution unit compared to {the} operation on background pixels (using MNIST data set). This activity results in power consumption and voltage drops inducing an increase in propagation delay of the TDC circuit that is adjacent to the BNN accelerator in the FPGA and that can be observed. TDC output voltage estimates are observed for each clock cycle of the first convolution layer. {The a}veraged estimations are represented using {their} HWs and, after applying filters to remove noise{,} a threshold is used to differentiate foreground {from} background pixels to finally reconstruct the shape of the input image. 
For their experiments, the authors considered a BNN pre-trained with MNIST data set and focus{ed} on the first convolution layer with a 28 $\times$ 28 gray scale image as the input and 64 kernels of size 3 $\times$ 3. An overview of the attack is illustrated in \figurename~\ref{fig:moini2021power}.
\textbf{Discussion and limitations:} The advantage of these attacks is that they can be performed without any physical proximity to the target device. However, using TDC-based sensor estimations requires them to be adjacent to the BNN accelerator (located in the next row of logical blocks). 
Moreover, this attack requires an important number of traces {captured using} the same image and kernel {for the} convolution{s, which might not be a very realistic scenario}. 
Therefore, it is required that the victim user requests the evaluation {of} the same image \textit{N} multiple times by the inference engine so that the adversary is able to collect enough traces (\textit{N} similar traces of the convolution of {the} same input and kernels). These traces are then averaged and filters are used in order to reduce noise. For evaluation purposes, {the authors performed the attack multiple times (from 100 to 6000 runs), with and without denoising, using the exact same input and kernel. For their experiments, the authors considered different {local} FPGA boards for accelerator implementation (Xilinx ZCU104 and VCU118) and different locations for the victim and attacker modules. {A similar experiment was run for Amazon AWS F1 FPGA instances that contain an Ultrascale+ device; in this case, the results are provided for 6000 runs}. Their experiments shown that {having the TDC sensors} adjacent to the victim module is required to perform this attack. Significant switching of other design components located in the proximity of the convolution unit might jeopardize the attack.
{The a}uthors evaluated their approach through normalised cross-correlation between the reconstructed images and the original ones and their methodology achieved a maximum normalised cross-correlation of 0.745 and 0.678 with the adjacent placement of TDC on the FPGA for ZCU104 and VCU118, respectively. The mean structural similarity index (MSSIM~\cite{wang2004image}) was also calculated. This index determines the similarity between two fixed-size windows of two images instead of evaluating the pixels individually. The MSSIM index is within the range of \{$-$1,1\}, {where} values close to 1 indicate a complete match while $-$1 indicates a complete mismatch. Within an 11{-}pixel sliding window, the authors' results displayed a MSSIM index around 0.8, which is compatible with {the} cross-correlation results.

\begin{figure}[H]
\resizebox{.9\linewidth}{!}{\includegraphics{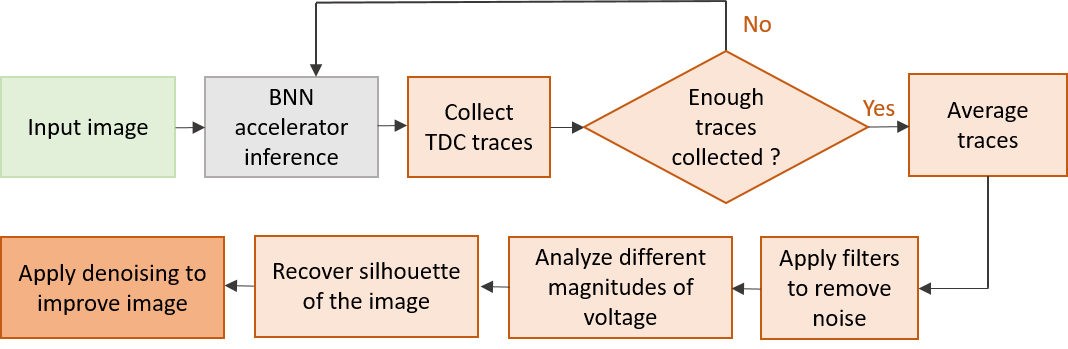}}
\caption{Overview of the attack proposed by Moini et al.~\cite{wei2018know} {(artwork reworked)}.}
\label{fig:moini2021power}
\end{figure}

\section{Current Countermeasures}
\label{sec:Current_countermeasures}

While the research field of SCA {attacks} against hardware implementations continues to gain interest, very few mitigation techniques have been proposed so far. 

{MaskedNet, by} Dubey et al.~\cite{dubey2020maskednet}, {re}present{s} the first implementation of {a} BNN on FPGA including power-based SCA defense mechanisms. Following the principle of masking-based techniques widely used in the cryptograph{ic} domain, the authors propose a solution adapted to the case of {D}NN{s} in order to reduce the induced costs.  
{The a}uthors start by implementing a baseline BNN FPGA-based design and a power-based attack using DPA. With this attack discussed in Section~\ref{sub:FPGA}, the authors show that an attacker can target the adder tree in order to extract the secret weights. Therefore, Dubey et al. designed a masking technique aimed at breaking the correlation between the summations computed at each state of the adder tree and the secret weights{, as} exploited to perform their attack.
The aim of masking is to make every intermediate computation independent of the secret data. The general idea is to split computations dependent on secret data into two (or more) \textit{randomized parts}. These parts are processed independently and reconstituted at the end in order to generate the expected result. As it is a defense mechanism at the level of the algorithm, it is therefore not specific to the implementation technology (FPGA or ASIC). The authors mainly proposed to mask two key components: the adder tree and the activation function.
For this, each input pixel $px_{i}$ is split into two parts{:} $p_{1} = r_{i}$ and $p_{2} = px_{i}-r_{i}$; $r$ being an 8-bit random number. $p_{1}$ and $p_{2}$ are independent of the input pixel value. Subsequently, the adder tree individually processes each part summation in such a manner that their combination will generate the original sum. 
In the same manner, the authors propose to mask the activation function of {the} BNN{, which} generates +1/$-$1 {when} the weighted summation is positive/negative. In the case of masking, the activation function will receive the summations corresponding to the two generated parts. The masked function then requires the determination of the sign of the summation of the two parts but without performing their actual summation. As the sign only depends on the MSB of the final summation{,} {the} authors designed a circuit that sequentially calculates and propagates the masked carry bits as in a ripple-carry generator.   
The authors highlighted that there is a correlation between the sign bit and the input. Therefore, a hiding technique was used for the sign bit computation in order to generate constant power consumption and to avoid information leakages.
For the experimental evaluation of the proposed defense, the authors perform first-order DPA-based SCA aimed at extracting the secret weight values on their baseline BNN FPGA-based design and the same BNN including the masked components. Results show that while only 200 traces were required for the attack to succeed on the unprotected network, the attack is still not successful after 100K traces in the protected design. As for the overhead induced by their defense, the authors evaluated an area increase in the number of FPGA LUTs (Look Up Tables), flip flops and BRAMs {(embedded Block RAM memories)} in their design of 2.7$\times$ , 1.7$\times$  and 1.3$\times${,} respectively, and an increase in inference latency of 2$\times$.
Finally, this defense is demonstrated to be effective for first-order SCA (DPA in their evaluation methodology){. H}owever{,} {the} authors note that splitting the secret data dependent inputs into more parts would enable the extension of the solution to be effective against higher-order SCA.

In addition to this first countermeasure that was completely implemented and evaluated, some literature contributions studying attacks have provided some insights of possible mitigation techniques. In order to mitigate power analysis, Batina et al.~\cite{batina2019csi} proposed to use a shuffle-based technique in order to modify the sequence of each neuron multiplication within a layer across different inference operations. They also highlight that while constant-time computation for instance of activation functions should be efficient against timing attacks, it might introduce significant performance overhead. 

Similarly, Maji et al.~\cite{maji2021leaky} proposed constant-time implementation of ReLu activation function{s} as well as the {fixed-point} representation of any generic floating point weight in order to avoid {the} timing leakage of ReLu f{u}nction{s} and floating point multipl{y} and accumulate operation{s}. However, the proposed defense was evaluated on ATmega328P micro-controller; the results shown a significant performance overhead ({the} time required to perform multiplication operation increased by 2$\times$).   

Moini et al.~\cite{moini2020remote,moini2021power} provided some ideas on how to mitigate input recovery attacks. Specifically, they propose to shuffle the pixel order of {the} convolution unit processing for each image input in order to increase the difficulty when trying to reconstruct the secret image. The overhead and efficiency of this technique remains to be evaluated.

\section{Discussion and Future Research Leads}
\label{sec:Discussion_and_future_research_leads}

In this section challenges of reviewed attacks are discussed. Then, some research directions for future work are provided. Finally, further SCA techniques out of the scope of this survey are introduced.

\subsection{Challenges of Studied Attacks}

Physical SC{A} attacks exploiting power and EM traces constitute an emergent research field that is rapidly gaining interest. These attacks have shown that they are a real and dangerous threat and that mitigation techniques are undoubtedly required. However{,} several limitations of {the} studied attacks can be identified.

\textbf{Reproducibility:} Very few work{s} reference (dedicated articles and/or source code) the {D}NN model and the implementation details for their evaluation. Experimental results{,} however{,} might significantly vary depending on implementation details. 

\textbf{Scalability/genericity:} As {this} is still an emergent research field, the majority of attack works aim at providing a proof of concept for the feasibility of SCA attacks as well as their potential effect. However, the scalability and genericity, for instance from a different physical target or a more complex and realistic data set{,} are not often considered. {Moreover,} some attacks are specific to a certain physical target or to a certain design choice.

\textbf{Metrics for evaluation:} Several metrics have been proposed for evaluating the efficiency of attacks including {the} accuracy of {the} recovered information, {the} portion of {the} recovered information and {the} similarity of results for victim and recovered networks (see Table~\ref{Classification_of_attacks_according_to_each_feature.}). We can note that some metrics serve for evaluation purposes when showing a proof-of-concept but cannot be used in an actual attack. This is, for example, the case of the comparison on the accuracy of {the} recovered information and, for instance, of recovered weight values. Comparing them to the victim {D}NN, the actual values are not possible as this information will not be accessible to the attacker.  
Moreover, when looking at comparing attacks or at studying their feasibility on {more} realistic scenarios, it is necessary to be able to evaluate their cost and complexity. The number of traces required for performing an attack are usually provided. However, very few works consider the overhead in terms of memory storage or recovery techniques computation effort. Similarly, as mitigation techniques emerge, common metrics for evaluating the resilience of defenses and their cost will be of great matter.

\subsection{Future Research Leads}

The first few defenses that have been implemented have shown a very significant overhead in terms of latency and area (approximately 2$\times$). This overhead represents an important challenge, especially for low-power devices in the IoT context. This motivates pursuing efforts on the design of further mitigation techniques an{d} on their optimization. With their emergence, attacks targeting protected {D}NN implementations are also expected to be studied in the near future.
After machine learning used for enhancing SCA and machine learning implementations being the target of SCA {attacks}, machine learning-enhanced SCA against machine learning implementations represent a promising future research direction. Fo{r} example, {an} interesting work~\cite{xiang2020open} {already} mixes both machine learning attacks (adversarial in this case) and SCA {attacks} against {D}NN implementations. This work has been discussed i{n} Section~\ref{sec:Physical_SCA_on_NN}.
Finally, current literature works target CNN{s/DNNs} as they have been widely studied and deployed in low-cost processors and FPGAs. However, other networks such as Recurrent {or} Spiking Neural Networks (RNN and SNN) are rapidly gaining interest. Studying their secure implementation is also a{n} interesting research direction. 

\subsection{Out of the Scope Attacks}

Finally, other {side-channels and} SC{A} techniques, which out of the scope of this review{,} have been successfully applied against {D}NNs. Several endeavors have exploited timing SC{A attacks} against {D}NNs running on different implementation platforms, software and hardware. Timing measurements are usually accessible through the analysis of cache access patter{n}s~\cite{yan2020cache}, which is a hardware resource shared between the attacker and victim application. However, the observation of other physical measurements including power or EM traces can also provide information on {computation time}~\cite{batina2019csi}. These attacks have been widely studied against $\times$86 processor-based {D}NN implementations.
Gongye et al.~\cite{gongye2020reverse}, for example, have successfully recovered the network parameters of a{n} MLP model {from a} software implementation on a $\times$86 processor, assuming that the attacker knows the model layout (number of layers and neurons in each layer). 
Cache-based SCA techniques, such as the well known Flush+Reload~\cite{yarom2014flush+}, have also been used against {D}NN software implementations in order to recover model architectures~\cite{yan2020cache,hong2018security}. A close work~\cite{gongye2020new} {has} recover{ed} the network parameters on a CNN model implemented in a $\times$86 processor through persistent cache monitoring. Contrary to the vast majority of {existing} literature attacks, the target model is trained with real medical data. 
The observation and analysis of off-chip memory access patterns have also been exploited for deducing {D}NN architecture layout{s} and parameters~\cite{hua2018reverse}. Other works~\cite{dong2019floating} have performed timing attacks using a timing model specifically {adapted} to micro-controllers in order to recover the network input data (data to be classified). Logical SC{A} attacks against {D}NN{s} as well as other attacks including trojan and fault injection and algorithmic-based attacks {are also} being studied~\cite{chabanne2021side}.  

\section{Summary}
\label{sec:Summary}

In this paper a survey of physical SCA {attacks} against embedded {D}NN implementations has been presented. {The} focus has {been placed} {on} Power-based and EM-based attacks. This work {has} shown that while there is an emergent and important research field on attacks, it is not easy to compare {existing works} as they follow different threat models, assumptions, targets and evaluation metrics. Moreover, several challenges have been identified, such as attack reproducibility, scalability and genericity. A taxonomy of attacks has been proposed and works have been classified according to their diverse features in order to provide a thorough overview of the existing techniques as well as their limitations.   
We have also presented the very few existing mitigation techniques that have been proposed in the literature so far. We have discussed their challenges and {we} hope to encourage to pursue further endeavors in this direction. We highlighted the necessity of following common attack scenarios and evaluation metrics for both attacks and defenses in order to easily compare different approaches.
Finally, we provided insights on future research leads.

\vspace{6pt} 




\funding{This research received no external funding.
}

\institutionalreview{Not applicable.
}

\informedconsent{Not applicable.

}

\dataavailability{Not applicable.
} 


\conflictsofinterest{The authors declare no conflicts of interest.} 

}\end{paracol}
\reftitle{References}

\end{document}